\documentstyle[times,prm1,epsf,floats,epsfig]{ias1}

\newcommand{\bm}[1]{\mbox{\boldmath $#1$}}
\newcommand{\beq}{\begin{equation}}
\newcommand{\eeq}{\end{equation}}

\begin{document}
\mark{{$\!\!\!\!\!\!\!\!\!\!\!\!\!\!\!\!\!\!\!\!\!\!\!\!\!\!\!\!\!\!$ QCD spin physics: 
status, and prospects for RHIC}
{$\!\!\!\!\!\!\!\!\!\!\!\!\!\!\!\!\!\!\!\!\!\!\!\!\!\!\!\!\!\!$ Werner Vogelsang}}

\title{QCD spin physics: status, and prospects for RHIC\footnote{Invited plenary talk
presented at the ``Workshop on High Energy Physics Phenomenology (WHEPP-8)'',
Indian Institute of Technology, Mumbai, January 5-16, 2004.}}

\author{Werner Vogelsang}
\address{RIKEN-BNL Research Center and BNL Nuclear Theory,
Brookhaven National Laboratory, Upton, NY 11973, U.S.A.}


\abstract{
\noindent We review some of the recent developments 
in QCD spin physics and highlight the spin
program now underway at RHIC.}

\maketitle

\vspace*{-6.8cm}
\begin{flushright}
BNL-NT-04/15 \\[1mm]
RBRC-416 
\end{flushright}
\vspace*{5.3cm}

\section{Introduction} 
For many years now, spin has played a very prominent
role in QCD. The field of QCD spin physics 
has been carried by the hugely successful
experimental program of polarized deeply-inelastic lepton-nucleon 
scattering (DIS), and by a simultaneous tremendous progress in theory. 
This talk summarizes some of the interesting new developments
in the past roughly two years. As we will see, there have yet 
again been exciting new data from polarized lepton-nucleon scattering, 
but also from the world's first polarized $pp$ collider, RHIC. 
There have been very significant advances in theory as well. 
It will not be possible to cover all developments. 
I will select those topics that may be of greatest interest
to the attendees of a high-energy physics phenomenology conference. 

\section{Nucleon helicity structure}
\subsection{What we have learned so far}
\begin{figure}[!h]
\vspace*{6.3cm}
\begin{center}
\includegraphics{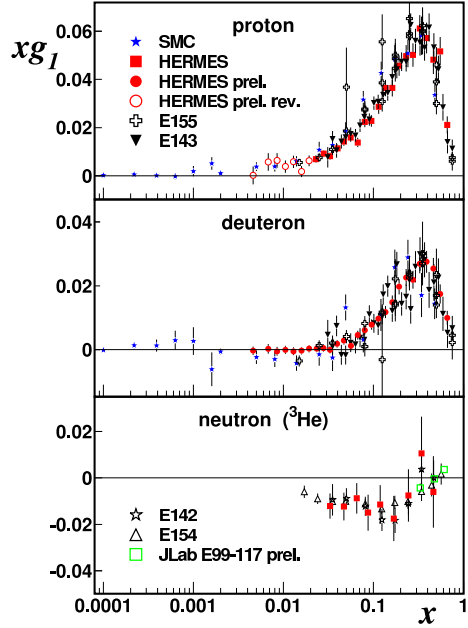}
\vspace*{2.5cm}
\caption[*]{Data on the spin structure function
$g_1$, as compiled and shown in \cite{US02}. \label{fig1}}
\end{center}
\end{figure}

Until a few years ago, polarized inclusive DIS played the dominant 
role in QCD spin physics \cite{hv}. 
At the center of attention was the nucleon's 
spin structure function $g_1(x,Q^2)$. Fig.~\ref{fig1} shows a recent
compilation~\cite{US02} of the world data on $g_1(x,Q^2)$. 
These data have provided much interesting information about the nucleon 
and QCD. For example, they have given direct access to the 
helicity-dependent parton distribution functions of the nucleon,
\begin{equation}
\Delta f(x,Q^2)=f^+ - f^- \;\;\;\;\;\;\; (f=q,\bar{q},g) \label{eq1}\; ,
\end{equation}
which count the numbers of partons with same helicity as the
nucleon, minus opposite.
Polarized DIS actually measures the combinations
$\Delta q+\Delta \bar{q}$. From $x\to 0$ extrapolation of the 
structure functions for proton and neutron targets
it has been possible to test and confirm the Bjorken sum 
rule \cite{bj}. Polarized DIS data, when combined with input from 
hadronic $\beta$ decays, have allowed to extract the -- unexpectedly 
small -- nucleon's axial charge $\sim\,\langle P|\bar{\psi} \,
\gamma^{\mu}\, \gamma^5 \, \psi |P\rangle$, which is identified
with the quark spin contribution to the 
nucleon spin \cite{hv}. 

\vspace*{-1.cm}
\subsection{Things we would like to know}
The results from polarized inclusive DIS have also
led us to identify the next important 
goals in our quest for understanding the spin structure
of the nucleon. The measurement of gluon
polarization $\Delta g=g^+-g^-$ rightly is a 
main emphasis at several experiments in spin physics 
today, since $\Delta g$ could be a major contributor to 
the nucleon spin. Also, more detailed understanding of polarized
quark distributions is clearly needed; for example, 
we would like to know about flavor symmetry breakings in
the polarized nucleon sea, details about strange quark
polarization, the relations to the $F,D$ values extracted
from baryon $\beta$ decays, and also about the small-$x$ and large-$x$
behavior of the densities. Again, these questions are being
addressed by current experiments. Finally, we would like
to find out how much orbital angular momentum quarks and 
gluons contribute to the nucleon spin. Ji showed \cite{ji} 
that their total angular momenta may be extracted 
from deeply-virtual Compton scattering, which
has sparked much experimental activity also in this area.

\subsection{Current experiments in high-energy spin physics} 
There are several fixed-target lepton-nucleon scattering 
experiments around the world with dedicated spin physics 
programs. I will mention those that play a role in this talk: 
{\sc Hermes} at DESY uses {\sc Hera}'s 27.5~GeV polarized electron beam 
on polarized targets. They have recently completed a run with a transversely 
polarized target. Semi-inclusive DIS (SIDIS) measurements are one 
particular strength of {\sc Hermes}. {\sc Compass} at CERN uses 
a 160~GeV polarized muon beam. A major emphasis is measuring gluon 
polarization. There is also a very large spin program at Jefferson Lab, 
involving several experiments. Large-$x$ structure functions 
and the DVCS reaction are just two of many objectives 
there. For the more distant future, there are plans to 
develop a polarized electron-proton {\it collider} at BNL, 
eRHIC \cite{eRHIC}.

A new milestone has been reached by the advent
of the first polarized proton-proton collider, RHIC at BNL. 
Two physics runs with polarized protons colliding 
at $\sqrt{s}=200$~GeV have been completed, and exciting results 
are emerging. We will see examples in this
talk. All components crucial for the initial phase of the spin 
program with beam polarization up to 50\% are in place \cite{bland}. 
This is true for the accelerator (polarized source, Siberian snakes, 
polarimetry by proton-Carbon and by $pp$ elastic scattering off a jet
target) as well as for the detectors. RHIC presently brings to collision 
55 bunches with a polarization pattern $\;\ldots++--++ 
\ldots\;$ in one ring and $\;\ldots+-+-+- \ldots\;$ in the other, 
which amounts to collisions with different spin combinations every 
212~nsec. Polarization has been maintained with a lifetime of about 
10 hours. There is still need for improvements in polarization and 
luminosity for future runs. The two larger RHIC experiments,
{\sc Phenix} and {\sc Star}, have dedicated spin programs focusing
on precise measurements of $\Delta g$, quark polarizations
by flavor, transverse-spin phenomena, and others.
A smaller experiment, {\sc Brahms}, investigates 
single-spin asymmetries. The pp2pp experiment studies
elastic $pp$ scattering.

\subsection{Accessing gluon polarization $\Delta g$}
As mentioned above, the measurement of $\Delta g$ is a 
main goal of several experiments. The gluon density
affects the $Q^2$-evolution of the structure function 
$g_1(x,Q^2)$, but the limited lever arm in $Q^2$ available 
so far has left $\Delta g$ virtually unconstrained \cite{grsv,bb,aac}.  
One way to access $\Delta g$ in lepton-nucleon scattering is therefore to 
look at a less inclusive final state that is particularly 
sensitive to gluons in the initial state. One channel, to be 
investigated by {\sc Compass} in particular \cite{compass}, 
is heavy-flavor production via the photon-gluon fusion process 
An alternative reaction is $ep\to h^+ h^- X $, where the
two hadrons in the final state have large transverse
momentum \cite{compass,bravar}. 

\begin{figure}[!b]
\vspace*{3.6cm}
\begin{center}
\includegraphics{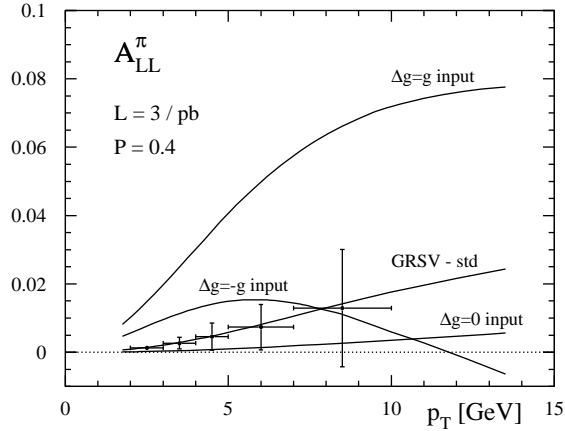}
\vspace*{2cm}
\caption[*]{NLO predictions \cite{jssv} for the spin asymmetry
in $pp\to \pi X$ at RHIC, for various $\Delta g$. The ``error bars''
are projections of the uncertainties that can be reached with 
40\% beam polarizations and 3/pb integrated luminosity. \label{fig2}}
\end{center}
\end{figure}

\begin{figure}[!b]
\vspace*{3.6cm}
\begin{center}
\includegraphics{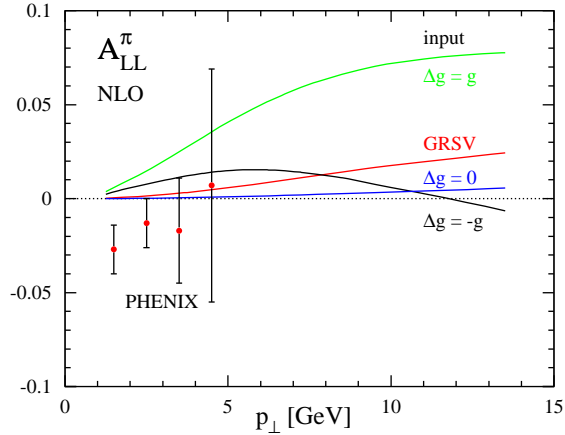}
\vspace*{2.2cm}
\caption[*]{{\sc Phenix} data \cite{dubna} for the spin asymmetry
$A_{\mathrm{LL}}^{\pi}$, along with the NLO predictions from 
the previous figure. \label{fig2a}}
\end{center}
\end{figure}

RHIC will likely dominate the measurements of $\Delta g$.
Several different processes will be investigated \cite{rhicrev} 
that are sensitive to gluon polarization: high-$p_{\perp}$ prompt photons 
$pp\to \gamma  X $, jet or hadron production $pp\to {\rm jet}X$, 
$pp\to h X$, and heavy-flavor production $pp\to (Q\bar{Q}) X$.
In addition, besides the current $\sqrt{s}=200$~GeV, also 
$\sqrt{s}=500$~GeV will be available at a later stage. All
this will allow to determine $\Delta g(x,Q^2)$ in various
regions of $x$, and at different scales. One can compare the
$\Delta g$ extracted in the various channels, and
hence check its universality implied by factorization
theorems. The latter state that cross sections at high 
$p_{\perp}$ (which implies large momentum transfer)
may be factorized into universal (process-independent)
long-distance pieces that contain the 
desired information on the (spin) structure of the nucleon, and 
short-distance parts that describe the hard interactions of the 
partons and are amenable to QCD perturbation theory (pQCD).
For example, for the reaction $pp\to \pi X$ one has:
\begin{eqnarray}
\label{eq:eq2}
d\Delta \sigma^{\pi} &=&\sum_{a,b,c}\, 
\Delta a \,\otimes \,\Delta b  \,\otimes\,
d\Delta \hat{\sigma}_{ab}^{c} \,\otimes \,D_c^{\pi}\; , 
\end{eqnarray}
where $\otimes$ denotes a convolution and where 
the sum is over all  contributing partonic channels $a+b\to c + X$, with
$d\Delta \hat{\sigma}_{ab}^{c}$ the associated spin-dependent 
partonic cross section. The $\Delta a,\Delta b \;(
a,b=q,\bar{q},g)$ are the polarized parton densities, and
the transition of parton $c$ into the observed $\pi^0$
is described by the (spin-independent) fragmentation function
$D_c^{\pi}$. We emphasize that all tools are in place
now for treating the spin reactions relevant at RHIC to
next-to-leading order (NLO) pQCD \cite{jssv,jsvj,vn,sv}. 
NLO corrections significantly improve
the theoretical framework; it is known from experience 
with the unpolarized case that the corrections are 
indispensable in order to arrive at quantitative predictions
for hadronic cross sections. For instance, the
dependence on factorization and renormalization scales in the
calculation is much reduced when going to NLO. Therefore, only with 
knowledge of the NLO corrections will one be able to
extract $\Delta g$ reliably. Figure~\ref{fig2} shows NLO 
predictions \cite{jssv} for the double-longitudinal spin 
asymmetry $A_{\mathrm{LL}}=d\Delta \sigma/d\sigma$ 
for the reaction $pp\to \pi X$ at RHIC, using various different 
currently allowed parameterizations \cite{grsv} of $\Delta g(x,Q^2)$. It
also shows the statistical errors bars expected for
a measurement by {\sc Phenix} under the rather conservative
assumptions of 40\% beam polarizations and 3/pb integrated 
luminosity. Such numbers are targeted for the early RHIC runs. 
Recently, first results for $A_{\mathrm{LL}}$ in $pp\to \pi X$ 
have indeed been reported by {\sc Phenix} \cite{dubna}, albeit
obtained with lower polarization and luminosity. The results are
shown in Fig.~\ref{fig2a}, along with the theoretical 
predictions that were already displayed in Fig.~\ref{fig2}. 
Interestingly, the data are consistent with a significant 
(up to a few per cent) negative asymmetry in the region $p_\perp 
\sim 1 \div 4$~GeV, contrary to all predictions shown in the 
figure. Even though the experimental uncertainties are still large
and leave room for a different behavior of $A_{\mathrm{LL}}^{\pi}$,
the new data give motivation to entertain the unexpected
possibility of $A_{\mathrm{LL}}^{\pi}$ being negative.
As it turns out \cite{jksv}, within pQCD at leading power,
there is a lower bound on the asymmetry of about $-10^{-3}$. 

To demonstrate this, we consider the LO cross section integrated
over all pion rapidities $\eta$ and take Mel\-lin 
moments in $x_T^2=4p_{\perp}^2/S$ of the cross section Eq.~(\ref{eq:eq2}):
\begin{equation} \label{doublemom}
\Delta\sigma^{\pi} (N) \equiv
\int_0^1 dx_T^2 \left( x_T^2 \right)^{N-1} 
\frac{p_{\perp}^3 d\Delta\sigma^{\pi}}{d p_{\perp}} \; .
\end{equation}
One finds:
\begin{equation} \label{crosec2}
\Delta\sigma^{\pi} (N) = \sum_{a,b,c} \,\Delta a^{N+1}\,
\Delta b^{N+1}\,\Delta
\hat{\sigma}_{ab}^{c,N}\,D_c^{\pi,2N+3}\; ,
\end{equation}
where the $\Delta\hat{\sigma}_{ab}^{c,N}$ are the $\hat{x}_T^2$-moments 
of the partonic cross sections and, as usual,
$ f^N\equiv \int_0^1 dx\, x^{N-1} f(x)$
for the parton distribution and fragmentation functions. 
Explicitly, the dependence on the moments $\Delta g^N$ of the
polarized gluon density is
\begin{equation} \label{quad1}
\Delta\sigma^{\pi} (N) =
\left(\Delta g^{N+1} \right)^2 {\cal A}^N + 
2 \Delta g^{N+1} {\cal B}^N + {\cal C}^N \; .
\end{equation}
Here, ${\cal A}^N$ represents the contributions from $gg\to gg$ and
$gg\to q\bar{q}$, ${\cal B}^N$ the ones from $qg\to qg$, and
${\cal C}^N$ those from the (anti)quark scatterings.  

Being a quadratic form in $\Delta g^{N+1}$, $\Delta\sigma^{\pi} (N)$
possesses an extremum, given by the condition \cite{jksv}
\begin{equation} \label{dgmin}
{\cal A}^N \Delta g^{N+1}  = -{\cal B}^N \; .
\end{equation}
The coefficient ${\cal A}^N$ is positive, and
Eq. (\ref{dgmin}) describes a minimum of $\Delta\sigma^{\pi} (N)$,
with value
\begin{equation} \label{crsecmin}
\Delta\sigma^{\pi} (N) \Big|_{\mathrm{min}} =
 -\left({\cal B}^N \right)^2/{\cal A}^N + {\cal C}^N \; .
\end{equation} 
It is then straightforward to perform a numerical Mellin
inversion of this minimal cross section.
The minimal asymmetry resulting from this exercise is negative indeed, 
but very small: in the range $p_\perp \sim 1 \div 4$~GeV
its absolute value does not exceed $10^{-3}$. The $\Delta g$
in Eq.~(\ref{dgmin}) that minimizes the asymmetry
has a node and is small, except at large $x$ \cite{jksv}.

Even though some approximations have been made in deriving the bound
in Eq.~(\ref{crsecmin}), it does exhibit the basic 
difficulty with a sizable negative $A_{\mathrm{LL}}^{\pi}$ at moderate
$p_\perp$: the fact that the cross section
is a quadratic form in $\Delta g$ effectively means that
it is bounded from below. Effects like NLO corrections,
choice of scales, and realistic range of rapidity may be
thoroughly addressed in a ``global'' NLO analysis of the data, taking
into account the results from polarized DIS as well. Such an analysis
has been performed in \cite{jksv}, and it confirms the findings
of the simple example above.
\begin{figure}[!b]
\vspace*{7.2cm}
\includegraphics{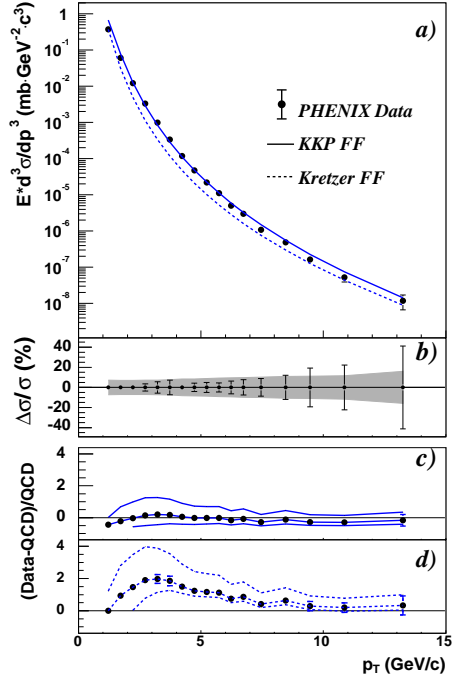}
\vspace*{2.4cm}
\caption[*]{{\sc Phenix} data \cite{phenixpi0} for the unpolarized
$pp\to \pi^0 X$ cross section at RHIC, compared to 
NLO calculations \cite{jssv}. 
The plot has been taken from \cite{phenixpi0}. \label{fig2b}}
\end{figure}

What should one conclude if future, more precise, data
will indeed confirm a sizable negative $A_{\mathrm{LL}}^{\pi}$?
Corrections to Eq.~(\ref{eq:eq2}) as such are down by
inverse powers of $1/p_{\perp}$. Since $p_\perp$ is not too large, 
such power-suppressed contributions might well be significant. 
On the other hand, comparisons of unpolarized $\pi^0$ spectra 
measured at RHIC with NLO QCD calculations do not exhibit 
any compelling trace of non-leading power effects even down to fairly 
low $p_\perp \gtrsim 1\ {\rm GeV}$, within the uncertainties of the
calculation. This is shown in Fig.~\ref{fig2b}. Clearly,
such results provide confidence that the theoretical 
hard scattering framework used for Figs.~\ref{fig2},\ref{fig2a}  
is indeed adequate. It is conceivable that the spin-dependent cross section 
with its fairly tedious cancelations has larger power-suppressed 
contributions than the unpolarized one. 

\subsection{Further information on quark polarizations}
As mentioned earlier, inclusive DIS via photon exchange 
only gives access to the combinations $\Delta q+\Delta 
\bar{q}$. There are at least two ways to distinguish between 
quark and antiquark polarizations, and also to achieve a 
flavor separation. Semi-inclusive measurements in DIS 
are one possibility, explored by SMC \cite{smc} and, more recently
and with higher precision, by {\sc Hermes} \cite{hermesdq}. 
\begin{figure}[!b]
\vspace*{5.4cm}
\includegraphics{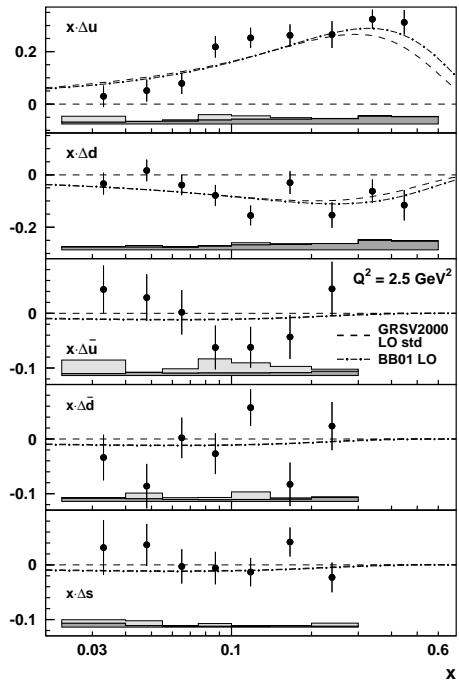}
\vspace*{3.7cm}
\caption[*]{Recent {\sc Hermes} results \cite{hermesdq} 
for the quark and antiquark
polarizations extracted from semi-inclusive DIS.  \label{fig3}}
\end{figure}
One detects a hadron in the final state, so that instead of $\Delta q+
\Delta \bar{q}$ the polarized DIS cross section becomes 
sensitive to $\;\Delta q(x)\,D_q^h(z) + \Delta  \bar{q}(x)\,
D_{\bar{q}}^h (z)\;$, for a given quark flavor. Here, the $D_i^h(z)$ are 
fragmentation functions, with $z=E^h/\nu$. Fig.~\ref{fig3} shows the 
latest results on the flavor separation by {\sc Hermes} 
\cite{hermesdq}, obtained from their LO Monte-Carlo code 
based ``purity'' analysis. Within the still fairly large 
uncertainties, they are not inconsistent with the large negative 
polarization of $\Delta\bar{u}=\Delta\bar{d}=\Delta \bar{s}$ in 
the sea that has been implemented in many determinations of polarized 
parton distributions from inclusive DIS data \cite{grsv,bb} (see curves 
in Fig.~\ref{fig3}). On the other hand, there is no evidence either 
for a large negative strange quark polarization. For the region 
$0.023<x<0.3$, the extracted $\Delta s$ integrates \cite{hermesdq} to 
the value $+0.03\pm 0.03 \,\mathrm{(stat.)}\,\pm 0.01 \,\mathrm{(sys.)}$,
while analyses of inclusive DIS typically 
prefer an integral of about $-0.025$. 
There is much theory activity currently on SIDIS, focusing also
on possible systematic improvements to the analysis method employed 
in \cite{hermesdq}, among them NLO corrections, target fragmentation, 
and higher twist contributions \cite{sidis}. At RHIC \cite{craigie83}
one will use $W^{\pm}$ production to determine $\Delta q,\Delta \bar{q}$,
making use of parity-violation. Figure~\ref{figw} 
shows the expected precision with which it will be possible to 
determine the light quark and antiquark polarizations.
Comparisons of such data taken at much higher 
scales with those from SIDIS will be extremely interesting. 
\begin{figure}[!h]
\vspace*{8cm}
\includegraphics{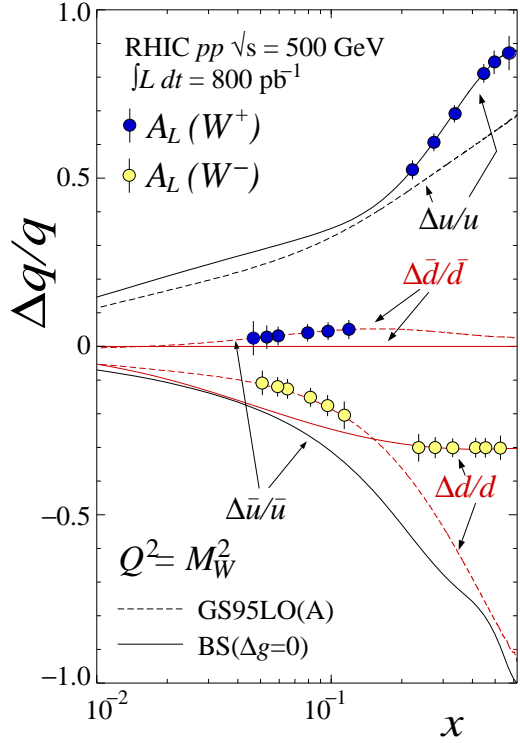}
\vspace*{28mm}
\caption[*]{Expected sensitivity \cite{rhicrev,craigie83} 
for the flavor decomposition
of quark and anti-quark polarizations at RHIC. \label{figw}}
\end{figure}

\begin{figure}[!h]
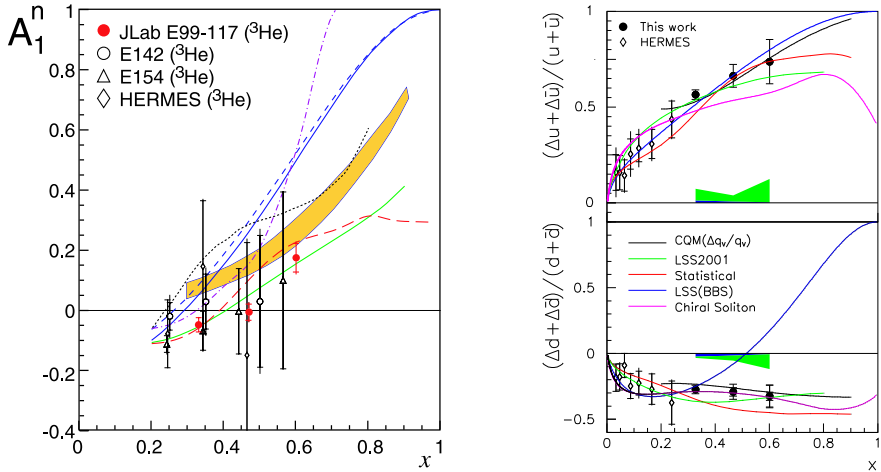

\vspace*{6.8cm}
\includegraphics{res_a1_zein.epsi}
\includegraphics{newjlab.epsi}
\vspace*{4mm}
\caption[*]{Left: Recent data on $A_1^n$ from the E99-117 experiment
\cite{e99117}. Right: extracted polarization asymmetries for $u+\bar{u}$
and $d+\bar{d}$. For more details and references on the various
model predictions, see \cite{e99117}. \label{fig4}}
\end{figure}

New interesting information on the polarized quark densities has also recently
been obtained at high $x$. The Hall A collaboration at JLab has 
published their data for the neutron asymmetry $A_1^n$ \cite{e99117}, 
shown in Fig.~\ref{fig4}. The new
data points show a clear trend for $A_1^n$ to turn positive
at large $x$. Such data are valuable because the
valence region is a particularly useful testing ground for
models of nucleon structure. Fig.~\ref{fig4} also
shows the extracted valence polarization asymmetries. The data
are consistent with constituent quark models \cite{cqm}
predicting $\Delta d/d\to -1/3$ at large $x$, while 
``hadron helicity conservation'' predictions based on perturbative 
QCD and the neglect of quark orbital angular
momentum \cite{hhc} give $\Delta d/d\to 1$ and tend to deviate from 
the data, unless the convergence to 1 sets in very late.

\section{Transverse-spin phenomena}
\vspace*{-0.2cm}
\subsection{Transversity} 
\begin{figure}[!h]
\vspace*{6.8cm}
\includegraphics{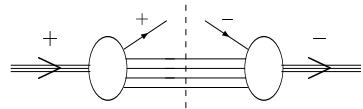}
\vspace*{-4.8cm}
\caption[*]{Transversity in helicity basis. \label{fig5}}
\end{figure}

Besides the unpolarized and the helicity-dependent
densities, there is a third set of twist-2 parton 
distributions, transversity \cite{rs}. In analogy with Eq.~(\ref{eq1})
they measure the net number (parallel minus antiparallel) 
of partons with transverse polarization in a transversely
polarized nucleon:
\begin{equation}
\delta f(x,Q^2)=f^{\uparrow} - f^{\downarrow} \label{eq2}\; .
\end{equation}
In a helicity basis, one finds \cite{rs} that transversity 
corresponds to a helicity-flip structure, as shown in Fig.~\ref{fig5}.
This precludes a gluon transversity distribution at 
leading twist. It also makes transversity
a probe of chiral symmetry breaking in QCD \cite{collins}: 
perturbative-QCD interactions preserve chirality,
and so the helicity flip required to make transversity non-zero must
primarily come from soft non-perturbative interactions for which 
chiral symmetry is broken \cite{collins}. 

\begin{figure}[!h]
\vspace*{-0.4cm}
\includegraphics{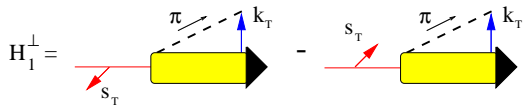}
\vspace*{2.6cm}
\caption[*]{The Collins function. \label{fig6}}
\end{figure}

Measurements of transversity are not straightforward. Again
the fact that perturbative interactions in the Standard
Model do not change chirality (or, for massless quarks, heli\-ci\-ty) 
means that inclusive DIS is not useful. Collins, however, 
showed \cite{coll93} that properties of fragmentation 
might be exploited to obtain a ``transversity polarimeter'': 
a pion produced in fragmentation will have some transverse 
momentum with respect to the momentum of the transversely polarized 
fragmenting parent quark. There may then be a correlation
of the form $\;i\vec{S}_T \cdot  (\vec{P}_{\pi} \times \vec{k}_{\perp})$. 
The fragmentation function associated with this correlation
is the Collins function. The phase is required by time-reversal 
invariance. The situation is depicted in Fig.~\ref{fig6}. 
The 
Collins function would make a {\it leading-power} \cite{coll93}
contribution to the single-spin asymmetry $A_{\perp}$ in the 
reaction $\;ep^{\uparrow}
\to e\pi X$:
\begin{equation} 
A_{\perp}\propto|\vec{S}_T|
\sin(\phi+\phi_S)\sum_q\,e_q^2
\delta q(x)H_1^{\perp,q}(z) \; ,\label{eq3}
\end{equation}
where $\phi$ ($\phi_S$) is the angle between the lepton plane
and the $(\gamma^* \pi)$ plane (and the transverse target spin).
We note that very recently a proof for the factorization formula 
for SIDIS at small transverse momentum was presented \cite{jmy}. 
As is evident from Eq.~(\ref{eq3}), the asymmetry would
allow access to transversity if the Collins functions are
non-vanishing. A few years ago, {\sc Hermes} measured the asymmetry 
for a longitudinally polarized target \cite{hermessp}. 
For finite $Q$, the target spin then has a transverse component
$\propto M/Q$ relative to the direction of the virtual photon, 
and the effect may still be there, even though it is now only one
of several ``higher twist'' contributions \cite{abko}. 

\subsection{The Sivers function} 
\begin{figure}[!t]
\vspace*{3.3cm}
\begin{center}
\includegraphics{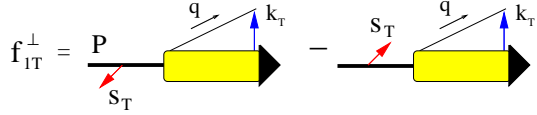}
\vspace*{-0.8cm}
\caption[*]{The Sivers function. \label{fig7}}
\vspace*{-0.7cm}
\end{center}
\end{figure}

If ``intrinsic'' transverse momentum in the fragmentation
process plays a crucial role in the asymmetry for 
$\;ep^{\uparrow} \to e\pi X$, 
a natural question is whether $k_{\perp}$ in the initial 
state can be relevant as well. Sivers suggested \cite{sivers}
that the $k_{\perp}$ distribution of a quark in a transversely
polarized hadron could have an azimuthal asymmetry, 
$\,\vec{S}_T \cdot  (\vec{P} \times \vec{k}_{\perp})$,
as shown in Fig.~\ref{fig7}. 
There is a qualitative difference between the Collins and 
Sivers functions, however. While phases will always arise in
strong interaction final-state fragmentation, one does not 
expect 
them from initial (stable) hadrons, and the Sivers
function appears to be ruled out by time-reversal invariance
of QCD \cite{coll93}. Until recently,
it was therefore widely believed that origins of single-spin 
asymmetries as in $\;ep^{\uparrow}\to e\pi X$ and other
reactions were more likely to be found in final-state 
fragmentation effects than in initial state parton distributions. 
However, then came a model calculation \cite{bbs} that
found a leading-power asymmetry in $\,ep^{\uparrow}\to e\pi X$
not associated with the Collins effect. It was subsequently 
realized \cite{coll02,bjy,bmp} that the calculation of \cite{bbs}
could also be regarded as a model for the Sivers effect. 
It turned out that the original time-reversal argument against 
the Sivers function is invalidated by the presence of the
Wilson lines in the operators defining the parton density.
These are required by gauge invariance and had been neglected
in \cite{coll93}. Under time reversal, however, future-pointing 
Wilson lines turn into past-pointing ones, which changes the 
time reversal properties of the Sivers function and allows
it to be non-vanishing. Now, for a ``standard'', $k_{\perp}$-integrated, 
parton density the gauge link contribution is unity in the $A^+=0$ 
gauge, so one may wonder how it can be relevant
for the Sivers function. The point, however, is that 
for the case of $k_{\perp}$-dependent parton densities, a gauge link
survives even in the light-cone gauge, in a transverse direction
at light-cone component $\xi^-=\infty$ \cite{bjy,bmp}. Thus, time 
reversal indeed does not imply that the Sivers function vanishes. 
The same is true for a function describing transversity in an 
unpolarized hadron \cite{boer}. It is intriguing that these new 
results are based entirely on the Wilson lines in QCD.  
Another aspect to the physics importance of the Sivers function is the 
fact that it arises as an interference of wave functions 
with angular momenta $J_z=\pm 1/2$ and hence contains information
on parton orbital angular momentum \cite{bbs,Ji}. 

\subsection{Implications for phenomenology}
If the Sivers function is non-vanishing, it will for example 
make a leading-power contribution to $\;ep^{\uparrow}\to e\pi X$,
of the form
\begin{equation} 
A_{\perp}\propto|\vec{S}_T|
\sin(\phi-\phi_S)\;\sum_q\,e_q^2\;
f_{1T}^{\perp ,q}(x)\;D_q^{\pi}(z)  \; .\label{eq4}
\end{equation}
This is in competition with the Collins function contribution,
Eq.~(\ref{eq3}); however, the azimuthal angular dependence
is discernibly different. {\sc Hermes} has recently completed 
an analysis of their data obtained in a run
with transverse target polarization, and preliminary results have
been presented, indicating contributions from both the
Collins and the Sivers effects \cite{hermtr}. A detailed study
\cite{schweitzer} suggests the surprising feature 
that the flavor-non-favored 
Collins functions appear to be equally important as the favored ones. 
{\sc Compass}, on the other hand, recently reported results for the
Collins asymmetries from a deuteron target, 
that are consistent with zero, within statistics \cite{comptr}.
We note that the Collins function may also
be determined separately from an azimuthal asymmetry in
$e^+e^-$ annihilation \cite{dbee}. It was pointed out 
\cite{coll02,bjy,bmp} that comparisons of DIS and the
Drell-Yan process will be particularly interesting:
from the properties of the Wilson lines it follows
that the Sivers functions relevant in DIS and in the 
Drell-Yan process have opposite sign, violating universality
of the distribution functions. This process dependence is a 
unique prediction of QCD. It is entirely calculable and
awaits experimental testing. For work on the process 
(in)dependence of the Collins function, see \cite{bmp,metz};
recent model calculations of the function in the context
of the gauge links may be found in \cite{schweitzer,models}.

A single-spin asymmetry in $p\,p$ scattering was identified 
recently \cite{bv} that also belongs to the class of 
``leading-power'' observables and may give access to Sivers
functions. The reaction considered was the inclusive 
production of jet pairs, $p\, p^\uparrow \to\mathrm{jet}_1 \, 
\mathrm{jet}_2 \, X$, for which the two jets are nearly back-to-back 
when projected into the plane perpendicular to the direction of the beams, 
which is equivalent to the jets being separated by nearly
$\Delta\phi\equiv\phi_{j_2}-\phi_{j_1}=\pi$ in azimuth.
This requirement makes the jet pairs sensitive to a 
small measured transverse momentum, and hence allows the single-spin
asymmetry for the process to be of leading power. 
\begin{figure}[b]
\vspace*{-4mm}
\begin{center}
\epsfig{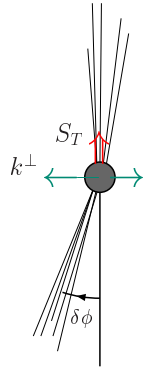}
\end{center}
\caption[*]{Asymmetric jet correlation. The proton beams run
perpendicular to the drawing. \label{fig:fig2}}
\vspace*{-0.5cm}
\end{figure}
The basic idea is very simple. The Sivers
function represents a correlation of the form ${\bm S}_T\cdot
({\bm P} \times {\bm k}^{\perp})$ between the transverse proton 
polarization vector, its momentum, and the transverse momentum of 
the parton relative to the proton direction. In other words, if 
there is a Sivers-type correlation then there will be a preference
for partons to have a component of intrinsic transverse momentum
to one side, perpendicular to both ${\bm S}_T$ and ${\bm P}$.
Suppose now for simplicity that one observes a jet in the direction
of the proton polarization vector, as shown in Fig.~\ref{fig:fig2}.
A ``left-right'' imbalance in ${\bm k}^{\perp}$ of the parton will then 
affect the $\Delta \phi$ distribution of jets nearly opposite to the 
first jet and give the cross section an asymmetric piece around 
$\Delta\phi=\pi$. The spin asymmetry $A_{\mathrm{N}}$ for this 
process  will extract this piece and give direct 
access to the Sivers function. In contrast to SIDIS,
it is rather sensitive to the nonvalence contributions to the 
Sivers effect, in particular the {\em gluon} Sivers function \cite{bv}. 

\begin{figure}[!t]
\begin{center}
\epsfig{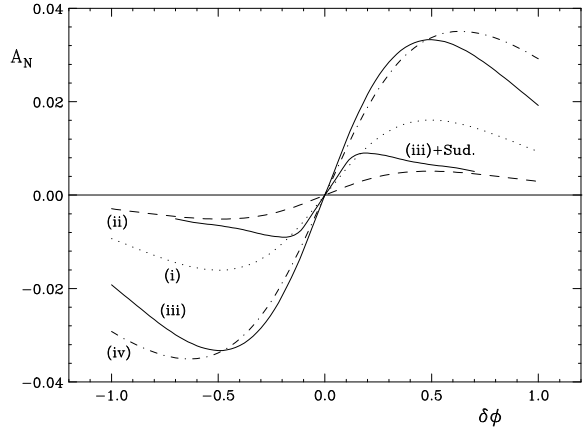}
\end{center}
\caption[*] {Predictions \cite{bv} for the spin asymmetry 
$A_{\mathrm{N}}$ for back-to-back dijet production at RHIC, 
for various different models for the gluon Sivers function. 
Note that $\delta \phi=\Delta \phi-\pi=0$, where
$\Delta \phi$ is the azimuthal separation of the jets.
The solid line marked as ``(iii)+Sud.'' shows the impact
of leading logarithmic Sudakov effects on the asymmetry curve
(iii). \label{fig:fig3}}
\end{figure}

Figure~\ref{fig:fig3} shows some predictions for the 
spin asymmetry in this reaction. Since nothing is known
about the size of the gluon Sivers function, some simple
models were made for it \cite{bv}, based on earlier studies 
of \cite{anselmino} for the valence quark Sivers 
distributions. For details, see \cite{bv}. 
One can see that sizable asymmetries are by all means possible. 
Near $\delta \phi=\Delta \phi-\pi=0$, however, 
gluon radiation is kinematically inhibited, 
and the standard cancelations of infrared singularities between 
virtual and real diagrams lead to large logarithmic remainders in the
partonic hard-scattering cross sections. It is possible to resum these 
Sudakov logarithms to all orders in $\alpha_s$. This was
done at the level of leading logarithms in \cite{bv}, for both the
numerator and the denominator of the asymmetry. 
As the analysis revealed, Sudakov effects lead to a significant 
suppression of the asymmetry, as  
is also visible from the solid lines in Figure~\ref{fig:fig3}.
This finding does not necessarily mean, however, 
that the asymmetry must be small,
since as we pointed out before, the gluon Sivers
function is entirely unknown and could well be larger
than in the models assumed for Figure~\ref{fig:fig3}.
In any case, any sign in experiment of a back-to-back asymmetry 
will be definitive evidence for the Sivers effect.  We note that 
for the back-to-back dijet distribution, the issue of whether 
or not factorization occurs still remains to be investigated. 

\begin{figure}[!h]
\vspace*{5.5cm}
\includegraphics{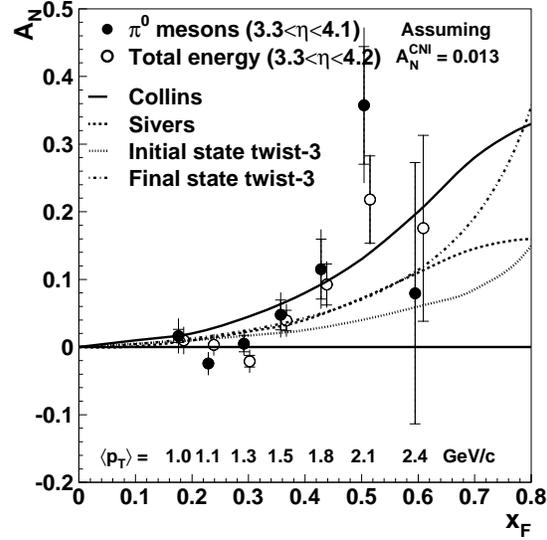}
\vspace*{26mm}
\caption[*]{Recent {\sc Star} results for the asymmetry $A_{\mathrm{N}}$ 
in $pp\to \pi^0X$ in the forward Feynman-$x_F$ 
region \cite{staran}. \label{fig9}}
\end{figure}

Originally, the Sivers function was proposed \cite{sivers} as a means
to understand and describe the significant single-spin 
asymmetries $A_{\mathrm{N}}$ observed \cite{anold} in $p^{\uparrow}p
\to \pi X$, with the pion at high $p_{\perp}$. These are {\em inclusive}
``left-right'' asymmetries and may be generated by the Sivers 
function from the effects of the quark intrinsic transverse momentum 
$k_{\perp}$ on the partonic hard-scattering which has a steep 
$p_{\perp}$ dependence. The resulting asymmetry $A_{\mathrm{N}}$ 
is then power-suppressed as $\sim\langle k_{\perp} \rangle /p_{\perp}$ in QCD, 
where $\langle k_{\perp} \rangle$ is an average intrinsic transverse
momentum. Similar effects may arise also from the Collins
function. Fits to the available $A_{\mathrm{N}}$ data have been  
performed \cite{dalesio}, assuming variously dominance of the Collins 
or the Sivers mechanisms. An exciting new development in the
field is that the {\sc Star} collaboration has presented the 
first data on $p^{\uparrow}p\to \pi X$ from RHIC \cite{staran}. 
The results are shown in Fig.~\ref{fig9}. As one can see,
a large $A_{\mathrm{N}}$ persists to these much higher energies.
Fig.~\ref{fig9} also shows predictions based on 
the Collins and the Sivers effects \cite{dalesio}, 
and on a formalism \cite{qs,koike} that
systematically treats the power-suppression of $A_{\mathrm{N}}$ in terms
of higher-twist parton correlation functions (for a connection
of the latter with the Sivers effect, see \cite{bmp}). 
The {\sc Star} data clearly give valuable information already now. For
the future, it will be important to extend the 
measurements to higher $p_{\perp}$ where the perturbative-QCD
framework underlying all calculations will become more 
reliable. We note that {\sc Star} has also measured the unpolarized
$pp\to \pi^0 X$ cross section in the same kinematic regime, 
which shows very good agreement with NLO pQCD calculations 
\cite{staran}. We note that the general consistency of RHIC
$pp\to \pi^0 X$ data with NLO pQCD results, already seen in Fig.~\ref{fig2b},
is in contrast to what was observed at lower energies in the 
fixed-target regime \cite{ftpion}.

\subsection{Two other developments}
It was recognized some time ago that certain Fourier transforms of 
generalized parton densities with respect to transverse momentum transfer 
give information on the position space distributions of partons 
in the nucleon \cite{ft}. For a transversely polarized nucleon, 
one then expects \cite{mb} a distortion of the parton distributions in
the transverse plane, which could provide an intuitive physical picture 
for the origins of single-spin asymmetries. 

Finally, {\it double}-transverse spin asymmetries 
$A_{\mathrm{TT}}$ in $pp$ scattering offer another possibility to access
transversity. Candidate processes are Drell-Yan, prompt photon, 
and jet production. Recently, the NLO corrections
to $p^{\uparrow}p^{\uparrow}\to \gamma X$ have been 
calculated \cite{msv}. The results show that $A_{\mathrm{TT}}$
is expected rather small at RHIC. It has also been proposed
\cite{agpax} to obtain transversity from the double-spin asymmetries
$A_{\mathrm{TT}}$ in Drell-Yan and $J/\psi$ production in 
possibly forthcoming polarized $\bar{p}p$ collisions at the GSI. 
An advantage here would be the fact that valence-valence scattering
is expected to dominate. On the other hand, the attainable energies 
may be too low for leading-power hard-scattering to clearly dominate.

\section*{Acknowledgments}
I am grateful to the organizers of WHEPP-8 for their 
invitation and hospitality, to D.\ Boer, B.\ J\"{a}ger, 
S.\ Kretzer, A.\ Mukherjee, and M.\ Stratmann for fruitful 
collaboration on various topics presented in this paper, 
and to M.\ Grosse-Perdekamp for helpful discussions. 
I thank RIKEN, BNL and the U.S. Department of Energy (contract number 
DE-AC02-98CH10886) for providing the facilities essential for 
the completion of this work.

\end{document}